\documentclass[runningheads]{llncs}
\usepackage{graphicx, amsmath, amssymb, cryptocode, tikz-cd, comment}

\newcommand{\F}{\mathbb{F}}
\newcommand{\FF}{\overline{\mathbb{F}}}

\newcommand{\N}{\mathbb{N}}

\newcommand{\Z}{\mathbb{Z}}
\newcommand{\G}{\mathcal{G}}

\begin{document}

\title{Undeniable signatures based on isogenies of supersingular hyperelliptic curves}
\titlerunning{Isogeny based undeniable signatures}
\author{Rams\`es Fern\`andez-Val\`encia\orcidID{0000-0002-8959-636X}} %
\authorrunning{R. Fern\`andez-Val\`encia} %
\institute{
Eurecat, Centre Tecnol\`ogic de Catalunya, IT Security Unit \\
Grup de Recerca en Nous Models de Ciberseguretat (2017 SGR 01239) \\
72 Bilbao Street, 08005 Barcelona, Catalonia \\
\email{ramses.fernandez@eurecat.org}}

\maketitle

\begin{abstract}
We present a proposal for an undeniable signature scheme based in supersingular hyperelliptic curves of genus $2$.
\end{abstract}


\section{Introduction}




An undeniable signature is a digital signature scheme which allows the signer to be selective to whom they allow to verify signatures. The scheme adds explicit signature repudiation, preventing a signer later refusing to verify a signature by omission; a situation that would devalue the signature in the eyes of the verifier.

Currently, there exists some concern about the definition of cryptographic algorithms able to resist attacks using a quantum computer. Among the techniques presumably able to lead to quantum-resistant cryptographic schemes, we find the use of isogenies of supersingular elliptic curves, which has proved to be an interesting solution in the definition of key exchanges \cite{Feo2015}, digital signatures \cite{Feo2018} and also has been explored in the definition of hash functions \cite{Charles2009}, \cite{Tachibana2017} and oblivious transfer protocols \cite{Barreto2018}.

As pointed out by \cite{Castryck2019}, there is a general awareness that many proposals, such as the isogeny-based Diffie-Hellman key exchange, should be generalized to principally polarized abelian surfaces. Among the motivations that support the research based on isogenies of higher genus curves we find the fact that the genus $2$ isogeny graph is much regular than the graph in the genus $1$ setting and this gives the chance to achieve similar security levels with less isogeny computations together with the opportunity to perform better security analysis. Furthermore, as noted in \cite{Takashima2018}, the number of $2$-isogenies between elliptic curves is three whereas abelian surfaces have fifteen $(2,2)$-isogenies, and this fact may improve the security of schemes using supersingular hyperelliptic curves.

This paper follows the philosophy in \cite{Castryck2019} and presents a proposal for a genus $2$ undeniable digital signature scheme based on the proposal in \cite{Jao2014} and using the techniques developed in \cite{Flynn2019}.

Concerning the structure of the document: the introduction is followed by the basics about the required mathematics (jacobians and Richelot isogenies, essentially) and undeniable signatures. Section 3 is devoted to the genus $2$ signature scheme, where we define the set-up, the key generation process and the signature procedure. The paper is closed with Section 4, devoted to the computationally complex problems involved in the security of the scheme, together with a few remarks on the correctness of our proposal.
\section{Fundamentals}\label{background}

This section is devoted to the mathematical requirements for the correct understanding of the forthcoming sections of this paper. The main reference for abelian varieties is \cite{Milne1986}, whereas \cite{Flynn2019} is a very good reference for the study of the structure of the isogeny graph in the genus-2 framework.

\subsection{Mathematical background} 

A Richelot isogeny is a $(2,2)$-isogeny between jacobians $J_H$ of genus $2$ curves $H$, that is: its kernel is isomorphic, as a group, to $\F_2\otimes \F_2$ and is maximal isotropic with respect to the $2$-Weil pairing.

Those familiar with isogeny graphs know that $j$-invariants play a central role in the definition of the graph. In the genus $2$ context the role played by the $j$-invariant is replaced with the so called $G_2$-invariants \cite{Cardona2002}. The $G_2$-invariants of a genus $2$ curve are absolute invariants that characterize the isomorphism class of the curve. Let $H$ be a genus $2$ curve and let $J_2$, $J_4$, $J_6$, $J_8$ and $J_{10}$ denote the associated Igusa invariants, then the $G_2$-invariants are defined as follows:
\begin{equation}
\mathcal{G} = (g_1, g_2, g_3) := \left( \frac{J_2^5}{J_{10}}, \frac{J^3_2J_4}{J_{10}}, \frac{J_2^2J_6}{J_{10}} \right).
\end{equation}

It is known (Chapter V, Section 13 \cite{Milne1986}) that an ample divisor $\mathcal{L}$ of an abelian variety $A$ defines an isogeny $\varphi_{\mathcal{L}}:A \to \hat{A}$ called a polarization of $A$. If $\varphi_{\mathcal{L}}$ is an isomorphism, then the polarization is called principal, and $A$ is a principally polarized abelian variety. The degree of a polarization is its degree as an isogeny.

The following result, allows us to work with jacobians of hyperelliptic curves of genus $2$:

\begin{theorem}[Theorem 1 \cite{Flynn2019}]
Given a prime $p$ and a finite field $\F_p$. If $A$ is a principally polarized abelian surface over $\FF_p$, then:
\begin{enumerate}
\item $A\cong J_H$, where $J_H$ denotes the jacobian of some smooth (hyperelliptic) genus $2$ curve $H$, or
\item $A \cong E_1 \times E_2$ for some elliptic curves $E_1$ and $E_2$. 
\end{enumerate}
\end{theorem}

\begin{proof}
The proof of (1) follows from Theorem 4 \cite{Oort1973} whereas (2) is a direct consequence of Theorem 3.1 \cite{Gonzalez2004}. 
\end{proof}

Another key ingredient is Proposition 1 \cite{Flynn2019} which proves the fact that isogenies with isotropic kernels preserve principal polarizations, what motivates using Richelot isogenies.

\begin{proposition}[Proposition 1 \cite{Flynn2019}]
Let $H$ be a hyperelliptic curve of genus $2$ over $\F_{p^n}$. Let $K$ be a finite, proper and $\F_{p^n}$-rational subgroup of $J_H(\F_{p^n})$. There exists a principally polarized abelian surface $A$ over $\F_{p^n}$ and an isogeny $\varphi: J_H\to A$ with kernel generated by $K$ if, and only if, $K$ is a maximal $m$-isotropic subgroup of $J_H[m]$ for some positive integer $m$.
\end{proposition}

\begin{proof}
The existence of $A$ follows immediately from Theorem 10.1 (Chapter VII, Section 10 \cite{Milne1986}). In order to prove that it is in fact a principally polarized abelian surface, one defines a polarization $\mu = [\deg(\varphi)]\circ \lambda$ on $J_H$, which is equipped with a principal polarization $\lambda$. One gets a polarization on $J_H/K$ of degree $1$ using Theorem 16.8 and Remark 16.9 (Chapter V, Section 16 \cite{Milne1986}).
\end{proof}

\subsection{Undeniable signatures}

We follow Section 4.1 \cite{Jao2014} in order to define an undeniable digital signature scheme as a tuple of algorithms
\begin{center}
{\bf UDS} = ({\bf KeyGen}, {\bf Sign}, {\bf Check}, {\bf CON}, {\bf DIS}),
\end{center}

where the role of {\bf CON} is for the signer to prove to the verifier that the signature is valid, whereas {\bf DIS} allows a valid signer to prove to the verifier that the signature received is not valid.

The scheme must be both unforgeable and invisible, where unforgeability is described with the following game between a challenger and an adversary $\mathfrak{A}$:
\begin{enumerate}
\item The challenger generates a pair of keys $(pbk, pvk)$ and provides $\mathfrak{A}$ with $pbk$.

\item Given some $q_s\in\N$ and for $i = 1,\dots, q_s$, $\mathfrak{A}$ queries {\bf Sign} adaptively with a message $\mu_i$ in order to obtain a signature $\sigma_i$.\label{sign}

\item $\mathfrak{A}$ outputs a forgery $(\mu^\star, \sigma^\star)$.
\end{enumerate}

The adversary $\mathfrak{A}$ is allowed to submit pairs $(\mu_j, \sigma_j)$ to {\bf Check} in step \ref{sign}, which proceed as follows:

\begin{enumerate}
\item If $(\mu_j, \sigma_j)$ is a valid pair, then {\bf Check} outputs a bit $b = 1$ and proceeds with the execution of {\bf CON}.

\item Otherwise {\bf Check} proceeds with the execution of {\bf DIS}. 
\end{enumerate}

The adversary $\mathfrak{A}$ succeeds in producing a strong forgery if $(\mu^\star, \sigma^\star)$ is valid and not among the pair generated during the queries in step \ref{sign}. The signature scheme is strongly unforgeable if the probability of $\mathfrak{A}$ succeeding in producing a strong forgery is negligible for any probabilistic polynomial time adversary $\mathfrak{A}$.

Concerning invisibility, it is defined through the following game between a challenger and an adversary:
\begin{enumerate}
\item The challenger generates a pair $(pbk, pvk)$ of keys and provides $\mathfrak{A}$ with $pbk$.

\item The adversary $\mathfrak{A}$ is allowed to issue, adaptively, a series of signing queries $\mu_i$ to {\bf Sign} and receive signatures $\sigma_i$. \label{sign_blind}

\item At some point, $\mathfrak{A}$ takes a message $\mu^\star$ and sends it to the challenger.

\item The challenger takes a random bit $b$. If $b = 1$, the he computes the real signature $\sigma^\star$ for $\mu^\star$ using $pvk$. Otherwise he computes a fake signature $\sigma^\dagger$ for $\mu^\star$ and sends it to $\mathfrak{A}$.

\item The adversary $\mathfrak{A}$ issues some more signing queries. \label{sign_again_blind}

\item At the end of the game, $\mathfrak{A}$ outputs a guess $b'$. 
\end{enumerate}

The adversary $\mathfrak{A}$ is allowed to submit pairs $(\mu_j, \sigma_j)$ to {\bf Check} adaptively in steps \ref{sign_blind} and \ref{sign_again_blind}, but it is not allowed to submit the challenge $(\mu^\star, \sigma^\star)$ to {\bf Check} in step \ref{sign_again_blind}. Further, $\mathfrak{A}$ is not allowed to submit $\mu^\star$ to {\bf Sign}. 

The signature scheme is invisible if no probabilistic polynomial time adversary $\mathfrak{A}$ has non-negligible advantage in this game.

For an undeniable signature to be secure, it is required to satisfy both unforgeability and invisibility. Furthermore, both {\bf CON} and {\bf DIS} must be complete, sound and zero-knowledge.

\section{Genus 2 undeniable digital signature scheme}

Here we define our undeniable digital signature scheme for supersingular hyperelliptic curves. The proposal relies heavily in the proposal made by Jao and Soukharev \cite{Jao2014} using the techniques used by the author in \cite{Fernandez2019}.

\subsection{Set-up}

Let us consider a prime of the form $p = l_A^{e_A}l_M^{e_M}l_C^{e_C}f \pm 1$ where $l_A,l_M,l_C$ and $f$ are integers and $f$ is small. Fix a supersingular hyperelliptic curve of genus $2$ which can be found by thinking it as the double cover of a supersingular elliptic curve of genus $1$. We then use a random sequence of Richelot isogenies to get a random principally polarized supersingular abelian surface. We also consider generating sets $\{A_1, A_2, A_3, A_4\}$, $\{M_1, M_2, M_3, M_4\}$ and $\{C_1, C_2, C_3, C_4\}$ for $J_H[l_A^{e_A}]$, $J_H[l_M^{e_M}]$ and $J_H[l_C^{e_C}]$, respectively. Finally, we take a hash function $\mathcal{H}: \{0,1\}^\star \to \Z^{12}$.

\subsection{Key generation}

The signer takes parameters $a_1,\dots, a_{12}\in \Z/l^{e_A}_A\Z$ following the techniques described in Section 3.2 \cite{Flynn2019} and computes an isogeny $\varphi_A: J_H \to J_A$, with kernel
\begin{equation}
K_A := \left\langle \sum_{i=1}^4[a_i]A_i, \sum_{i=5}^8[a_i]A_{i-4}, \sum_{i=9}^{12}[a_i]A_{i-8} \right\rangle,
\end{equation}

together with $\{\varphi_A(C_i)\}_{i = 1,2,3,4}$ and the invariants $\G(J_A)$. Then:
\begin{itemize}
\item $pbk := \left( \G(J_A), \{\varphi_A(C_i)\}_{i=1,2,3,4} \right)$.
\item $pvk := \{a_i\}_{i=1,\dots, 12}$.
\end{itemize}

\subsection{Signing}

Let $\mathcal{M}$ be the space of admitted messages. For a given message $\mu\in\mathcal{M}$ the signer computes the hash value $\mathcal{H}(\mu) = (h_1, \dots, h_{12})$. The signer also computes
\begin{equation}
K_M := \left\langle \sum_{i=1}^4[h_i]M_i, \sum_{i=5}^8[h_i]M_{i-4}, \sum_{i=9}^{12}[h_i]M_{i-8} \right\rangle,
\end{equation}

together with the following isogenies and points:
\begin{itemize}
\item $\varphi_M: J_H\to J_M = J_H/\langle K_M \rangle$.  
\item $\varphi^M_{AM}: J_M \to J_{AM} = J_M/\langle \varphi_M(K_A) \rangle$.
\item $\varphi^A_{AM}: J_A \to J_{AM} = J_A/\langle \varphi_A(K_M) \rangle$.
\item $\{\varphi^M_{AM}(\varphi_M(C_i))\}_{i = 1,2,3,4}$.
\item $\mathcal{G}(J_{AM})$.
\end{itemize}

The signature is defined as $\sigma := (\mathcal{G}(J_{AM}), \{\varphi^M_{AM}(\varphi_M(C_i))\}_{i = 1,2,3,4})$.

The following diagram provides a global view of the above calculations and of the signing procedure:

\begin{equation}\label{commutative}
\begin{tikzcd}
J_H \arrow[r, "\varphi_A"]\arrow[d, "\varphi_M" ] & J_A \arrow[d, "\varphi^A_{AM}"] \\
J_M \arrow[r, "\varphi^M_{AM}"] & J_{AM}
\end{tikzcd}
\end{equation}

\subsection{The algorithms CON and DIS}
The algorithms {\bf CON} and {\bf DIS} are required when we need to prove that a received signature is valid or when the signer is required to prove that a signature is not valid, respectively. Both algorithms have a common part:
\begin{enumerate}
\item The signer takes parameters $c_1,\dots, c_{12}\in \Z/l^{e_C}_C\Z$, according to Section 3.2 \cite{Flynn2019}, and computes:
	\begin{enumerate}
	\item[i.] The isogeny $\varphi_C:J_H\to J_C$ whose kernel is given by
	\begin{equation}
	K_C := \left\langle \sum_{i=1}^4[c_i]C_i, \sum_{i=5}^8[c_i]C_{i-4}, \sum_{i=9}^{12}[c_i]C_{i-8} \right\rangle.
	\end{equation}
	\item[ii.] The isogeny $\varphi^M_{MC}: J_M \to J_{MC}$ whose kernel is given 		by
	\begin{equation}\label{jmc}
	K_{MC} = \left\langle \sum_{i=1}^4[c_i]\varphi_M(C_i), \sum_{i=5}^8[c_i]\varphi_M(C_{i-4}), \sum_{i=9}^{12}[c_i]\varphi_M(C_{i-8}) \right\rangle.
	\end{equation}
	We observe that 
	\begin{equation}
	J_{MC}\cong \frac{J_M}{K_{MC}} = \frac{J_M}{\langle \varphi_M(K_C) \rangle}  \cong \frac{J_C}{\langle \varphi_C(K_M) \rangle} 
	\end{equation}		
	\item[iii.] $J_{AC} = \frac{J_A}{\langle \varphi_A(K_C) \rangle} \cong \frac{J_C}{\langle \varphi_C(K_A) \rangle}$.
	\item[iv.] The isogeny $\varphi^C_{MC}: J_C \to J_{MC}$ whose kernel is given by $K_{MC}$ and	
	\item[v.] $J_{AMC} = \frac{J_{MC}}{\langle \varphi^C_{MC}(K_A) \rangle}$.
	\end{enumerate}

\item The signer outputs $\G(J_C), \G(J_{AC}), \G(J_{MC}), \G(J_{AMC})$ and $\ker\left(\varphi^C_{MC}\right)$ as the commitment.

\item The verifier takes a random bit $b\in\{0,1\}$.
\end{enumerate}
 
We now describe the algorithm {\bf CON}:
\begin{itemize}
\item If $b = 0$: the signer sends $\ker(\varphi_C)$. Using $pbk$, the verifier computes $\ker\left( \varphi^A_{AC} \right)$. Using $\ker(\varphi_M)$, the verifier computes $\varphi^M_{MC}$. Using the digital signature, the verifier may compute $\varphi^{AM}_{AMC}$ and, with $	\ker(\varphi_C)$, the verifier can compute $\varphi^C_{MC}$ in order to check the commitment.

\item If $b = 1$: the signer sends $\ker\left( \varphi^C_{AC} \right)$ and the verifier computes $\varphi^{MC}_{AMC}$ together with $\varphi^{AC}_{AMC}$ and checks that each $\varphi^C_{AC}, \varphi^{MC}_{AMC}$ and $\varphi^{AC}_{AMC}$ maps between the curves specified in the commitment.
\end{itemize}

Concerning {\bf DIS}, let $J_F$ be a falsified version of $J_{AM}$ and $F_i$, for $i=1,2,3,4$, be the falsified points corresponding to $\varphi^M_{AM}(\varphi_M(C_i))$. Then:
\begin{itemize}
\item If $b = 0$: the signer outputs $\ker(\varphi_C)$ and the verifier computes $\varphi_C$, $\varphi^M_{MC}$, $\varphi^A_{AC}$ and $\varphi_F: J_F\to J_{FC}$, whose kernel is given by:
\[
K_F = \left\langle \sum_{i=1}^4[c_i]F_i, \sum_{i=5}^8[c_i]F_{i-4}, \sum_{i=9}^{12}[c_i]F_{i-8} \right\rangle.
\]

\item If $b = 1$: the signer outputs $\ker\left( \varphi^C_{AC} \right)$ and the verifier computes $\varphi^{AC}_{AMC}$ and $\varphi^{MC}_{AMC}$ and checks that these isogenies map to $J_{AMC}$.
\end{itemize}
\section{Complexity and security}

It is important to observe that the security analysis of the genus $1$ problems extends to our setting easily and so, those interested in further details are invited to read \cite{Galbraith2018}. 

Let us consider a prime of the form $p = l^{e_A}_Al^{e_M}_Cl^{e_C}_C f \pm 1$, a supersingular hyperelliptic curve $H$ over $\F_{p^2}$ and bases $\{A_1, A_2, A_3, A_4\}$, $\{M_1, M_2, M_3, M_4\}$ and $\{C_1, C_2, C_3, C_4\}$ for $J_H[l^{e_A}_A]$, $J_H[l^{e_M}_M]$ and $J_H[l^{e_C}_C]$ respectively.

The following problems are assumed to be infeasible in a quantum setting:

\begin{problem}[DSSI: Decisional supersingular isogeny problem]\label{dssi}
Let $H'$ be another supersingular hyperelliptic curve over $\F_{p^2}$ with genus $2$. Decide whether $J_{H'}$ is $(l^{e_A}_A, l^{e_A}_A)$-isogenous to $J_H$.
\end{problem} 

\begin{problem}[CSSI: Computational supersingular isogeny problem]\label{cssi}
Let $H_A$ be a supersingular hyperelliptic curve over $\F_{p^2}$ with genus $2$. Let $\varphi_A: J_H \to J_A$ be an isogeny whose kernel is generated by

\begin{equation}
K_A = \left\langle \sum_{i=1}^4[a_i]A_i, \sum_{i=5}^8[a_i]A_{i-4}, \sum_{i=9}^{12}[a_i]A_{i-8} \right\rangle
\end{equation}

for some $a_i\in\F_{2^n}$. Given $J_A$ and $\{\varphi_A(M_j)\}_{j=1,\dots, 4}$, find generators for $K_A$.
\end{problem}

\begin{problem}[SSCDH: Supersingular computational Diffie-Hellman]\label{sscdh}
Let us consider $\varphi_A: J_H \to J_A$ an isogeny whose kernel is

\begin{equation}
K_A = \left\langle \sum_{i=1}^4[a_i]A_i, \sum_{i=5}^8[a_i]A_{i-4}, \sum_{i=9}^{12}[a_i]A_{i-8} \right\rangle,
\end{equation}

for some $a_i\in\F_{2^n}$ and let $\varphi_M:J_H\to J_M$ whose kernel is

\begin{equation}
K_M = \left\langle \sum_{i=1}^4[m_i]M_i, \sum_{i=5}^8[m_i]M_{i-4}, \sum_{i=9}^{12}[m_i]M_{i-8} \right\rangle
\end{equation}

for some $m_i\in\F_{l^{e_M}_M}$. Given $\{\varphi_A(M_j), \varphi_M(A_j)\}_{j=1,\dots, 4}$ and the jacobians $J_A, J_M$, find the set $\mathcal{G}$ associated to $\frac{J_H}{(K_A, K_M)}$.
\end{problem}

\begin{problem}[SSDDH: Supersingular decision Diffie-Hellman]\label{ssddh}
Given a tuple sampled with probability $1/2$ from one of the following two distributions:
\begin{enumerate}
\item $(J_A, J_B, \{\varphi_A(M_i)\}_i, \{\varphi_M(A_i)\}_i, J_{AM})$, where $J_A$, $J_M$, $\{\varphi_A(M_i)\}_i$, $\{\varphi_M(A_i)\}_i$ are as in Problem \ref{sscdh} and $J_{AM} = J_H/\langle K_A, K_M \rangle.$

\item $(J_A, J_B, \{\varphi_A(M_i)\}_i, \{\varphi_M(A_i)\}_i, J_C)$, where $J_A$, $J_M$, $\{\varphi_A(M_i)\}_i$, $\{\varphi_M(A_i)\}_i$ are as in Problem \ref{sscdh} and $J_C = J_H/\langle K_{A'}, K_{M'} \rangle$, with 
\begin{align*}
K_{A'} =& \left\langle \sum_{i=1}^4[a'_i]A_i, \sum_{i=5}^8[a'_i]A_{i-4}, \sum_{i=9}^{12}[a'_i]A_{i-8} \right\rangle \\
K_{M'} =& \left\langle \sum_{i=1}^4[m'_i]M_i, \sum_{i=5}^8[m'_i]M_{i-4}, \sum_{i=9}^{12}[m'_i]M_{i-8} \right\rangle
\end{align*}
where $\{a'_i\}_i\in\Z/l^{e_A}_A$ and $\{m'_i\}_i\in\Z/l^{e_M}_M$ are chosen following the criteria in Section 3.2 \cite{Flynn2019}.
\end{enumerate}
determine from which distribution the tuple sampled.
\end{problem}

\begin{problem}[DSSP: Decisional supersingular product]
Given an $(l_A^{e_A},l_A^{e_A})$-isogeny $\varphi_A:J_0\to J_1$ and a tuple sampled with probability $1/2$ from one of the following two distributions:
\begin{enumerate}
\item $(J_1, J_2,\varphi')$ where the product $J_1\times J_2$ is chosen at random among those $(l_B^{e_B},l_B^{e_B})$-isogenous to $J_0 \times J_1$ and where $\varphi':J_1\to J_2$ is a random $(l_A^{e_A},l_A^{e_A})$-isogeny, and
\item $(J_1, J_2,\varphi')$ where $J_1$ is chosen at random among those surfaces with the same cardinality as $J_0$ and $\varphi':J_1\to J_2$ is a random $(l_A^{e_A},l_A^{e_A})$-isogeny,
\end{enumerate}
determine which distribution the tuple is sampled.
\end{problem}

\begin{problem}[MSSCDH: Modified supersingular computational Diffie-Hellman]\label{msscdh}
Keeping the notations used in problem \ref{ssddh}, given $J_A, J_M$ and $\ker(\varphi_M)$, determine whether $J_C = J_{AM}$. 
\end{problem}

\begin{problem}[MSSDDH: Modified supersingular decision Diffie-Hellman]\label{mssddh}
Keeping the notations used in problem \ref{ssddh}, given $J_A, J_M, J_C$ and $\ker(\varphi_M)$, determine $J_{AM}$.
\end{problem}

\begin{problem}[1MSSCDH: Modified supersingular computational Diffie-Hellman, one-sided version]\label{1msscdh}
Given $J_A, J_M$ and an oracle to solve problem \ref{msscdh} for any $J_A, J_{M'}, \ker(\varphi_{M'})$, where $J_{M'}\ncong J_M$, solve problem \ref{msscdh} for $J_A, J_M$ and $\ker(\varphi_M)$.
\end{problem}

\begin{problem}[1MSSDDH: Modified supersingular decisional Diffie-Hellman, one-sided version]
Given $J_A, J_M, J_C$ and an oracle to solve problem \ref{mssddh} for any $J_A, J_{M'}, \ker(\varphi_{M'})$ where $J_{M'}\ncong J_M$, solve problem \ref{mssddh} for $J_A, J_M, J_C$ and $\ker(\varphi_{M})$.
\end{problem}

\subsection{Remarks on security: completeness, soundness, zero-knowledge, unforgeability and invisibility}

Concerning the security of these algorithms, it is required to prove that both {\bf CON} and {\bf DIS} are complete, sound and zero-knowledge. We recall \cite{Goldreich2004_1} that a verification system is complete if some prover is able to convince the verifier of true statements, whereas soundness describes the ability of the verifier to detect false statements.

When it comes to prove completeness, soundness and zero-knowledge for {\bf CON} and {\bf DIS}, we may use the reasoning in \cite{Jao2014} having in mind that random integers are now replaced by scalars satisfying certain conditions that guarantee maximality and isotropicity. Furthermore, there is a subtlety concerning completeness for the {\bf DIS} protocol: if $J_F = J_{AM}$ contains subgroups $K_1, K_2$ such that $J_{AM}/K_1 = J_{AM}/K_2$, then $J_{AM}$ would be a branch point in the covering space of the jacobian of the modular curve $X_0(l^{e_C}_C)$, which is hyperelliptic in a few specific cases \cite{Ogg1974}. Therefore the chance of $J_{AM}$ being equal to such a jacobian is negligible.

In order to prove unforgeability and invisibility, we can use, again after adapting the methods for our genus-2 context, the reasonings of \cite{Jao2014}. 







\end{document}